\documentclass[%
 preprint,
 amsmath,
 amssymb,
 showpacs
]{revtex4-1}
\usepackage{float}
\usepackage{graphicx}
\usepackage{dcolumn}
\usepackage{bm}
\usepackage{bbold}
\usepackage[export]{adjustbox}
\usepackage{amsmath,amsfonts,mathtools}
\usepackage{dsfont}
\usepackage{amsmath}
\usepackage{amssymb}
\usepackage{makeidx}
\usepackage{graphicx}
\usepackage{color}

\newcommand{\hide}[1]{}

\newcommand{\lcase}{\left\{\begin{array}{ll}}
\newcommand{\rcase}{\end{array}\right.}
\newcommand{\ear}{\end{array}}
\newcommand{\bal}{\begin{align}}
\newcommand{\eal}{\end{align}}
\newcommand{\bma}{\begin{pmatrix}}
\newcommand{\ema}{\end{pmatrix}}
\newcommand{\beq}{\begin{equation}}
\newcommand{\eeq}{\end{equation}}
\newcommand{\bel}[1]{\begin{equation}\label{eq:#1}}
\newcommand{\eel}{\end{equation}}
\newcommand{\bea}{\begin{eqnarray}}
\newcommand{\eea}{\end{eqnarray}}
\newcommand{\beaNN}{\begin{eqnarray*}}
\newcommand{\eeaNN}{\end{eqnarray*}}

\newcounter{lecture}

\renewcommand{\hide}[1]{}

\newcommand{\hydroplus}{\text{H}_2^+}

\begin{document}


\title{Infinite-time surface flux for full-dimensional three-body breakup dynamics}
\author{Jinzhen Zhu}
\email{zhujinzhenlmu@gmail.com}
\affiliation{%
 Physics Department, Ludwig Maximilians Universit\"at, D-80333 Munich, Germany
}%
\affiliation{%
 Shanghai Artificial Intelligence Laboratory, 129 Longwen Road, Shanghai, China
}

\begin{abstract}
We derive an infinite-time surface-flux formulation for full-dimensional three-body breakup dynamics in intense laser fields.
The method is designed as a post-pulse extension of time-dependent surface flux (tSurff) calculations for systems with two asymptotic fragments, with helium double ionization and dissociative ionization of $\hydroplus$ as representative applications.
Standard tSurff calculations avoid projection on very large boxes, but the spectra still contain a field-free tail after the laser pulse; converging this tail by direct propagation can be expensive for slow particles, narrow resonances, and long-range Coulomb channels.
Here the post-pulse time integrals are rewritten as resolvents of the field-free one-particle ionic Hamiltonians and of the full field-free three-body Hamiltonian.
The resulting expressions separate the already available tSurff amplitudes from stationary correction terms that can be evaluated from saved wave functions in the inner and single-ionization regions.
The formulation gives a common theoretical structure for electron-electron breakup in helium and electron-nuclear breakup in $\hydroplus$, and it is compatible with the spectral decompositions and MPI-parallel workflow of the tRecX framework.
This provides a practical route to tSurff+iSurff calculations of correlated three-body spectra without long post-pulse propagation and without solving a large complex linear system independently for every final momentum.
\end{abstract}

\maketitle
\section{Introduction}
The correlated breakup of few-particle Coulomb systems in intense laser fields is a central problem in strong-field and attosecond physics.
For helium, double ionization directly probes electron-electron correlation in the continuum, while dissociative ionization of $\hydroplus$ requires the simultaneous description of electronic emission and nuclear motion beyond the Born-Oppenheimer approximation.
Both cases are genuine three-body breakup problems: the desired observables are differential in two asymptotic momenta, and long-range Coulomb interactions strongly influence convergence with respect to simulation time and volume.

Time-dependent surface flux methods have made such calculations feasible by replacing projection on very large boxes with the flux through channel surfaces.
The original tSurff formulation for one-electron emission \cite{Tao2012} and its two-electron extension \cite{Scrinzi2012,Zielinski2016} provide fully differential spectra from wave functions represented only in a finite numerical volume.
In previous applications, this strategy was used for double emission from helium \cite{Zhu2020} and for full-dimensional dissociative ionization of $\hydroplus$ \cite{Zhu2020b,Zhu2021}.
These calculations demonstrated that tSurff can extract correlated electron-electron and electron-nuclear spectra from realistic time-dependent wave packets.

A remaining limitation is the treatment of the field-free signal after the laser pulse.
In standard tSurff, the wave function must be propagated until the outgoing packet has crossed the analysis surfaces and the residual post-pulse flux is negligible.
This continuation can dominate the cost for slow fragments, long pulses, narrow spectral structures, and multidimensional Coulomb breakup channels.
For one-particle ionization, infinite-time surface flux methods remove this bottleneck by expressing the field-free tail through resolvents evaluated at the end of the pulse \cite{Morales2016}.
The same idea has recently been included in the tRecX/haCC framework, where iSurff is combined with the recursive discretization and parallel infrastructure of tRecX \cite{Chundayil2024}.

The available iSurff constructions are, however, primarily one-particle asymptotic formulations.
Full-dimensional tSurff calculations exist for two-electron helium and for electron-nuclear breakup in $\hydroplus$, but a general infinite-time continuation for correlated three-body breakup has not, to our knowledge, been given in a form directly usable for both systems.
The extension is nontrivial because the post-pulse correction contains independent one-fragment propagation as well as sequential flux terms: after one particle has reached an asymptotic region, the remaining particle can still propagate in an ionic or coupled three-body field before reaching its own surface.

In this work we derive a common tSurff+iSurff framework for three-body quantum dynamics, using helium and $\hydroplus$ as the guiding examples.
The derivation rewrites the infinite post-pulse time integrals as resolvents of the field-free one-particle ionic Hamiltonians and of the full field-free three-body Hamiltonian.
This converts the missing tail of the tSurff signal into stationary overlap problems evaluated from wave functions already produced in the $B$, $S$, and $\bar{S}$ regions.
In an implementation, these operations can be organized by spectral decompositions and energy-indexed files, making the expensive part parallel over final energies and channels.
The resulting method avoids direct propagation to very large times and avoids a separate large complex linear solve for every final momentum point.

The paper is organized as follows.
We first define the common three-body Hamiltonian and the channel partition used for helium and $\hydroplus$.
We then derive the post-pulse iSurff correction for one ordering of the two outgoing particles and obtain the exchanged contribution by symmetry.
Finally, we discuss how the resulting expressions can be implemented in the tRecX framework using saved tSurff wave functions and parallel spectral decompositions.

\section{Theoretical framework}
We consider a three-body Hamiltonian of the form
\begin{equation}\label{eq:HamiltonianDISec2}
 H(t)=H_{1}(\vec{r_1},t)\otimes \mathds{1} + \mathds{1} \otimes H_{2}(\vec{r_2},t) + V,
\end{equation}
where $H_1$ and $H_2$ contain the one-particle kinetic, binding, and laser-coupling terms and $V$ contains the residual two-particle interaction.
The corresponding Volkov wave functions for the continuum channels are defined by 
\begin{equation}
i\partial_t\chi_{\vec{k_1}}(\vec{r_1},t)=H_{V,1}\chi_{\vec{k_1}}(\vec{r_1},t),\quad i\partial_t\chi_{\vec{k_2}}(\vec{r_2},t)=H_{V,2}\chi_{\vec{k_2}}(\vec{r_2},t).
\end{equation}
For helium, the two single-particle Hamiltonians are identical,
\begin{equation}\label{eq:ionic}
\begin{split}
 H_1(\vec{r},t)=H_2(\vec{r},t)=-\frac{\Delta }{2}-i\vec{A}(t)\cdot\vec{\nabla}-\frac{2}{r}\\
  H_{V,1}(\vec{r},t)=H_{V,2}(\vec{r},t)=-\frac{\Delta }{2}-i\vec{A}(t)\cdot\vec{\nabla},
\end{split}
\end{equation}
and interaction potential
\begin{equation}
  V=V_{ee}=\frac{1}{|\vec{r_1}-\vec{r_2}|}.
\end{equation}
The field-free Volkov energies are 
\begin{equation}
  E_1=\frac{k_1^2}{2},\qquad E_2=\frac{k_2^2}{2}.
\end{equation}
We use $E_j$ for scalar channel energies and define $\hat{E}_j=E_j\mathds{1}_j$ as the corresponding multiplication operator on the Hilbert space of coordinate $\vec{r}_j$ when tensor-product notation is needed.
For the $\hydroplus$ system, the two coordinates represent nuclear and electronic motion, and the corresponding single-particle Hamiltonians are
\begin{equation}\label{eq:ionicH2Plus}
\begin{split}
  H_1(\vec{r_1},t)&=-\frac{\Delta }{4M}+\frac{1}{2r_1}\\
  H_2(\vec{r_2},t)&=-\frac{\Delta }{2m}-i\beta\vec{A}(t)\cdot\vec{\nabla}\\
  H_{V,1}(\vec{r_1},t)&=-\frac{\Delta }{4M}\\
  H_{V,2}(\vec{r_2},t)&=-\frac{\Delta }{2m}-i\beta\vec{A}(t)\cdot\vec{\nabla}
\end{split}
\end{equation}
and interaction potential
\begin{equation}
  V=V_{pe}=-\frac{1}{|\vec{r_1}-\vec{r_2}|}-\frac{1}{|\vec{r_1}+\vec{r_2}|},
\end{equation}
where $M=1836$ is the proton mass in atomic units, $m=2M/(2M+1)$ is the reduced electronic mass, and $\beta=(M+1)/M$ is the laser-coupling mass factor.
Here $\vec{r}_1$ denotes the internuclear coordinate and $\vec{r}_2$ denotes the electronic coordinate with respect to the nuclear center of mass.
The field-free Volkov energies are 
\begin{equation}
  E_1=\frac{k_1^2}{4M},\qquad E_2=\frac{k_2^2}{2m}.
\end{equation}
\par
The asymptotic scattering amplitude is
\begin{equation}\label{eq:twoEbk1k2Infty}
  b(\vec{k_1},\vec{k_2})=\lim_{T\to \infty} b(\vec{k_1},\vec{k_2},T),
\end{equation}
and the corresponding joint energy spectrum is
\begin{equation}
 P(\vec{k_1},\vec{k_2})=|b(\vec{k_1},\vec{k_2})|^2,
\end{equation}
where
\begin{equation}\label{eq:twoEbk1k2}
  b(\vec{k_1},\vec{k_2},T)=\langle\langle \chi_{\vec{k_1}}(\vec{r_1},T)\otimes \chi_{\vec{k_2}}(\vec{r_2},T) \left |\Theta_1(R_c) \Theta_2(R_c) \right | \psi(\vec{r_1},\vec{r_2},T) \rangle \rangle.
\end{equation}
The scattering amplitudes at a time $T$ can be derived using tSurff as
\begin{equation}\label{eq:integralAmplitudes}
 b(\vec{k_1},\vec{k_2},T)
 =\int_{-\infty}^{T}[F(\vec{k_1},\vec{k_2},t)+ \bar{F}(\vec{k_1},\vec{k_2},t)]dt,
\end{equation}
where
\begin{equation}\label{eq:contributionF}
\begin{split}
   F(\vec{k_1},\vec{k_2},t)
 =&\langle \chi_{\vec{k_2}}(\vec{r_2},t)|S_2|\varphi_{\vec{k_1}}(\vec{r_2},t)\rangle \\
   \bar{F}(\vec{k_1},\vec{k_2},t)
 =&\langle \chi_{\vec{k_1}}(\vec{r_1},t)|S_1|\varphi_{\vec{k_2}}(\vec{r_1},t)\rangle, 
\end{split}
\end{equation}
The surface commutators are
\begin{equation}
  S_1=[H_{V,1}(\vec{r_1},t),\Theta_{1}(R_c)],S_2=[H_{V,2}(\vec{r_2},t),\Theta_{2}(R_c)],
\end{equation}
the single-particle auxiliary functions obey
\begin{equation}
\begin{split}
  i\partial_t\varphi _{\vec{k_1}}(\vec{r_2},t)=&H_{2}(\vec{r_2},t)\varphi _{\vec{k_1}}(\vec{r_2},t)-C_{\vec{k_1}}(\vec{r_2},t)\\
  i\partial_t\varphi _{\vec{k_2}}(\vec{r_1},t)=&H_{1}(\vec{r_1},t)\varphi _{\vec{k_2}}(\vec{r_1},t)-C_{\vec{k_2}}(\vec{r_1},t)
\end{split}
\end{equation}
with source terms
\begin{equation}\label{eq:source1}
 C_{\vec{k_1} }(\vec{r_2},t)=\langle \chi_{\vec{k_1}}(\vec{r_1},t)\left|S_1\right|\psi(\vec{r_1},\vec{r_2},t) \rangle
\end{equation}
and
\begin{equation}\label{eq:source2}
 C_{\vec{k_2} }(\vec{r_1},t)=\langle \chi_{\vec{k_2}}(\vec{r_2},t)\left|S_2\right|\psi(\vec{r_1},\vec{r_2},t) \rangle
\end{equation}
\par
To extend the propagation time to infinity, we write the scattering amplitude as

\begin{equation}
\begin{split}
    b(\vec{k_1},\vec{k_2}, \infty )=& 
    b(\vec{k_1},\vec{k_2}, T )+\int_{T}^{\infty}[F(\vec{k_1},\vec{k_2},t)+ \bar{F}(\vec{k_1},\vec{k_2},t)]dt.
\end{split}
\end{equation}
The first term is the amplitude accumulated by the standard tSurff calculation. The remaining task is to compute the second, field-free, post-pulse contribution, where $T$ is chosen as the end of the pulse.
This second contribution is field-free, so all Hamiltonians involved are time independent.
For notational brevity, we shift the post-pulse starting time to $T=0$.
Because the two terms are related by exchange symmetry, it is sufficient to derive the contribution from $F(\vec{k_1},\vec{k_2},t)$.
We use the following abbreviations:
\begin{equation}
\begin{split}
\psi_{0}&=\psi(\vec{r_1},\vec{r_2},0)\\
\chi_{0,1}&=\chi_{\vec{k_1}}(\vec{r_1},0),\chi_{0,2}=\chi_{\vec{k_2}}(\vec{r_2},0)\\
\varphi_{0,1}&=\varphi_{\vec{k_2}}(\vec{r_1},0),\varphi_{0,2}=\varphi_{\vec{k_1}}(\vec{r_2},0)\\
\end{split}
\end{equation}
The solution of the auxiliary equation can be written as
\begin{equation}
  \varphi _{\vec{k_1}}(\vec{r_2},t)=\exp(-iH_{2}t)\varphi_{0,2}-\int_{0}^{t}\exp(-iH_{2}(t-\tau))C(\tau)d\tau
\end{equation}
where $H_{2}$ is time independent. In the following, $H_{ion}$ denotes the field-free one-particle Hamiltonian governing the remaining coordinate after the first particle has crossed its surface; for the present ordering this is $H_2$. Thus 
\begin{equation}
\begin{split}
&\int_{0}^{\infty}F(\vec{k_1},\vec{k_2},t)dt\\
  =&\int_{0}^{\infty}\left< \chi_{\vec{k_2}}(\vec{r_2},t)\left | S_2\right | \varphi _{\vec{k_1}}(\vec{r_2},t)\right>\\
  =&\int_{0}^{\infty}\left< \chi_{0,2}\exp(-iE_2t)\left | S_2\right | \exp(-iH_{2}t)\varphi_{0,2}
  -\int_{0}^{t}\exp(-iH_{2}(t-\tau))C(\tau)d\tau \right>dt\\
  =&\int_{0}^{\infty}\left< \chi_{0,2}\left| S_2 \exp[-i(H_{ion}-E_2)t] \right| \varphi_{0,2}\right>-
 \int_{0}^{\infty}\left< \chi_{0,2}\left| S_2 \right|\exp(iE_2t)\int_{0}^{t}\exp(-iH_{2}(t-\tau))C(\tau)d\tau \right>dt\\
 =&P_{2,1}-P_{2,2}
\end{split}
\end{equation}
\begin{equation}
  P_{2,1}=\int_{0}^{\infty}\left< \chi_{0,2}\left| S_2 \exp[-i(H_{ion}-E_2)t] \right| \varphi_{0,2}\right>dt
\end{equation}
\begin{equation}
  P_{2,2}=\int_{0}^{\infty}\left< \chi_{0,2}\left| S_2 \right|\exp(iE_2t)\int_{0}^{t}\exp(-iH_{2}(t-\tau))C(\tau)d\tau \right>dt
\end{equation}
\begin{equation}
\begin{split}
    P_{2,1} =& \lim_{\epsilon\to 0} \left< \chi_{0,2}\left| S_2\frac{1}{\hat{H}_{ion}-E_2\mathds{1}+i\epsilon}\right|\varphi_{0,2} \right>\\
    =&\lim_{\epsilon\to 0} \left< \chi_{0,2}\left| S_2\frac{1}{\hat{H}_{2}-E_2+i\epsilon}\right|\varphi_{0,2} \right>
\end{split}
\end{equation}

This expression can be evaluated by solving the stationary equation
\begin{equation}
  [\hat{H}_{ion}-E_2]\varphi'=\varphi_{0,2}
\end{equation}
where the outgoing prescription is supplied by the complex-scaled representation, followed by the final overlap.
\begin{equation}
\begin{split}
  P_{2,2}&=\lim_{T\to\infty}\int_{0}^{T}\left< \chi_{0,2}\left| S_2 \right|\exp(iE_2t)\int_{0}^{t}\exp(-iH_{2}(t-\tau))C(\tau,\vec{r_2})d\tau \right>dt\\
  &=\lim_{T\to\infty}\int_{0}^{T}\left< \chi_{0,2}\left| S_2 \right|\exp(iE_2t)\int_{0}^{t}\exp(-iH_{2}(t-\tau))\langle \chi_{\vec{k_1}}(\vec{r_1},\tau)\left|S_1\right|\psi(\vec{r_1},\vec{r_2},\tau) \rangle d\tau \right>\\
  &=\lim_{T\to\infty}\int_{0}^{T}dt\left< \chi_{0,2}\left| S_2 \right|\exp(iE_2t)\int_{0}^{t}d\tau\exp(-iH_{2}(t-\tau))\langle \chi_{0,1}\exp(-iE_1\tau)\left|S_1\right|\exp(-iH\tau)\psi_0 \rangle \right>\\
  &=\lim_{T\to\infty}\int_{0}^{T}d\tau\left< \chi_{0,2}\left| S_2 \right|\exp(iE_2t)\int_{\tau}^{T}dt \exp(-iH_{2}(t-\tau))\langle \chi_{0,1}\exp(-iE_1\tau)\left|S_1\right|\exp(-iH\tau)\psi_0 \rangle \right>\\
  &=\int_{0}^{\infty}d\tau\left< \chi_{0,2}\left| S_2 \right|\exp(iE_2t)\int_{\tau}^{\infty}dt \exp(-iH_{2}(t-\tau))\langle \chi_{0,1}\exp(-iE_1\tau)\left|S_1\right|\exp(-iH\tau)\psi_0 \rangle \right>\\
  &=\lim_{\epsilon_2\to 0}\int_{0}^{\infty}d\tau\left< \chi_{0,2}\left| S_2 \right|\frac{i\exp(iE_2\tau)}{\hat{H}_{2}-E_2+i\epsilon_2}  \langle \chi_{0,1}\exp(-iE_1\tau)\left|S_1\right|\exp(-iH\tau)\psi_0 \rangle \right>\\
  &=\lim_{\epsilon_2\to 0}\int_{0}^{\infty}d\tau\left< \chi_{0,1}\otimes \chi_{0,2}\left| \hat{S}_1 \hat{S}_2 \right|\frac{i}{\mathds{1}\otimes\hat{H}_{ion}-\mathds{1}\otimes\hat{E}_2+i\epsilon_2} \exp[-i(\hat{H}-\mathds{1}\otimes\hat{E}_2 -\hat{E}_1\otimes \mathds{1})\tau]\psi_0 \right>\\
  &=-\lim_{\epsilon_1,\epsilon_2\to 0}\left< \chi_{0,1}\otimes \chi_{0,2}\left| \hat{S}_1 \hat{S}_2 \right|\frac{1}{\mathds{1}\otimes\hat{H}_{ion}-\mathds{1}\otimes\hat{E}_2+i\epsilon_2}
  \frac{1}{\hat{H}-\mathds{1}\otimes\hat{E}_2 -\hat{E}_1\otimes \mathds{1}+i\epsilon_1}\psi_0 \right>
\end{split}
\end{equation}
In the fourth line of this derivation the integration order has been exchanged over the triangular domain $0\leq \tau\leq t\leq T$, giving $0\leq \tau\leq T$ and $\tau\leq t\leq T$ before the limit $T\to\infty$ is taken. In the following step, the remaining $t$-integral is understood with the shift $s=t-\tau$, so that the phase factor separates as $\exp(iE_2t)=\exp(iE_2\tau)\exp(iE_2s)$ and the $s$-integration gives the resolvent of $\hat{H}_2-E_2$.
This expression can be evaluated by the two stationary steps
\begin{equation}
  [\hat{H}-\mathds{1}\otimes\hat{E}_2 -\hat{E}_1\otimes \mathds{1}]\psi'=\psi_0
\end{equation}
and
\begin{equation}
  [\mathds{1}\otimes\hat{H}_{2}-\mathds{1}\otimes\hat{E}_2]\psi''=\psi'
\end{equation}
applied sequentially in the complex-scaled representation.
The formal $i\epsilon$ factors in the resolvents specify the outgoing limit. Following the iSurff strategy in tRecX \cite{Chundayil2024}, after irECS or complex scaling the relevant continuum eigenvalues are rotated into the lower complex plane, so the discretized inverse is evaluated directly at $\epsilon=0$, as in the one-particle iSurff implementation. The required operations can then be performed through spectral decompositions of $\hat{H}_{2}$ and of the full field-free matrix $\hat{H}$.
\par
To summarize, 
\begin{equation}\label{eq:total}
  \begin{split}
&\int_{0}^{\infty}F(\vec{k_1},\vec{k_2},t)dt=P_{2,1}-P_{2,2}\\
=&\lim_{\epsilon\to 0} \left< \chi_{0,2}\left| \hat{S}_2\frac{1}{\hat{H}_{ion}-E_2+i\epsilon}\right|\varphi_{0,2} \right>\\
+&\lim_{\epsilon_1,\epsilon_2\to 0}\left< \chi_{0,1}\otimes \chi_{0,2}\left| \hat{S}_1 \hat{S}_2 \right|\frac{1}{\mathds{1}\otimes\hat{H}_{2}-\mathds{1}\otimes\hat{E}_2+i\epsilon_2}
  \frac{1}{\hat{H}-\mathds{1}\otimes\hat{E}_2 -\hat{E}_1\otimes \mathds{1}+i\epsilon_1}\psi_0 \right>
\end{split}
\end{equation}
and similarly
\begin{equation}
  \begin{split}
&\int_{0}^{\infty}\bar{F}(\vec{k_1},\vec{k_2},t)dt=\\
=&\lim_{\epsilon'\to 0} \left< \chi_{0,1}\left| \hat{S}_1\frac{1}{\hat{H}_{ion}-E_1+i\epsilon'}\right|\varphi_{0,1} \right>\\
+&\lim_{\epsilon'_1,\epsilon'_2\to 0}\left< \chi_{0,1}\otimes \chi_{0,2}\left| \hat{S}_1 \hat{S}_2 \right|\frac{1}{\hat{H}_{1}\otimes\mathds{1}-\hat{E}_1\otimes\mathds{1}+i\epsilon'_2}
  \frac{1}{\hat{H}-\mathds{1}\otimes\hat{E}_2 -\hat{E}_1\otimes \mathds{1}+i\epsilon'_1}\psi_0 \right>
\end{split}
\end{equation}
\par
In practice, the above two formulas can be written as
\begin{equation}
  \begin{split}
\int_{0}^{\infty}F(\vec{k_1},\vec{k_2},t)dt
&=\lim_{\epsilon\to 0} \left< \chi_{0,2}\left| \hat{S}_2\frac{1}{\hat{H}_{ion}-E_2+i\epsilon}\right|\varphi_{0,2} + \psi'_{0,2}(E_1,E'_2)\delta_{E'_2,E_2} \right>\\
\psi'_{0,2}(E_1,E'_2)&=\lim_{\epsilon'_1\to 0}\left< \chi_{0,1}\left| \hat{S}_1 \right|
  \frac{1}{\hat{H}-\mathds{1}\otimes\hat{E}'_2 -\hat{E}_1\otimes \mathds{1}+i\epsilon'_1}\psi_0 \right>
\end{split}
\end{equation}
and
\begin{equation}
  \begin{split}
\int_{0}^{\infty}\bar{F}(\vec{k_1},\vec{k_2},t)dt
&=\lim_{\epsilon\to 0} \left< \chi_{0,1}\left| \hat{S}_1\frac{1}{\hat{H}_{ion}-E_1+i\epsilon}\right|\varphi_{0,1} + \psi'_{0,1}(E'_1,E_2)\delta_{E'_1,E_1} \right>\\
\psi'_{0,1}(E'_1,E_2)&=\lim_{\epsilon'_1\to 0}\left< \chi_{0,2}\left| \hat{S}_2 \right|
  \frac{1}{\hat{H}-\mathds{1}\otimes\hat{E}_2 -\hat{E}'_1\otimes \mathds{1}+i\epsilon'_1}\psi_0 \right>.
\end{split}
\end{equation}
In practice, the algorithm consists of the following steps:
\begin{itemize}
  \item Propagate in the $B$ region and save the wave function $\psi_0$ at the end of the pulse.
  \item Propagate in the $S$ and $\bar{S}$ regions, save $\varphi_{0,2}$ and $\varphi_{0,1}$, and compute the energy-indexed files $\psi'_{0,2}(E_1,E'_2)$ and $\psi'_{0,1}(E'_1,E_2)$.
  \item Compute the iSurff correction from overlaps with the saved wave functions.
\end{itemize}

\section{Implementation and scaling}
The formulas above are intended as a post-processing extension of a standard tSurff calculation rather than as a replacement for the time propagation during the laser pulse.
The time-dependent calculation first stores the end-of-pulse wave function in the inner $B$ region and the one-particle auxiliary wave functions in the $S$ and $\bar{S}$ regions.
After the pulse, the iSurff correction is evaluated by applying the field-free resolvents in Eqs.~\eqref{eq:total} and its exchanged counterpart.
In a spectral representation of $\hat{H}_{1}$, $\hat{H}_{2}$, and the field-free three-body Hamiltonian $\hat{H}$, these resolvents reduce to energy denominators with the usual outgoing-wave prescription $i\epsilon$.
The computation therefore separates naturally into independent energy and channel blocks.

This organization is important for the intended tRecX implementation.
The expensive spectral decompositions are performed once for the field-free operators, while the overlaps for different final momenta can be distributed independently across MPI ranks.
The method avoids extending the time propagation until all slow outgoing fragments have reached asymptotic distances.
It also avoids solving a new large complex linear system for every pair of final momenta; instead, the momentum dependence enters through overlaps with saved channel functions and through scalar resolvent denominators in the spectral representation.
The numerical validation of this strategy will be presented separately, but the present derivation fixes the objects that must be stored and the sequence of operations required by an implementation.

\begin{acknowledgments}
The author thanks Prof. Armin Scrinzi for bringing the iSurff implementation in tRecX to his attention and for helpful discussions on the status and possible extension of iSurff ideas to two-particle emission.
These discussions helped motivate the present derivation and the formulation of the open problem addressed here.
The author is also grateful for the suggestion to compare with related post-pulse analysis approaches in the literature.
\end{acknowledgments}

\bibliography{h2plus.bib}
\end{document}